\documentclass[%
reprint,
superscriptaddress,
amsmath,amssymb,
aps,
]{revtex4-2}
\usepackage[utf8]{inputenc}
\usepackage[english]{babel}
\usepackage{blindtext}
\usepackage{graphicx}   
\graphicspath{{figs/}}   
\usepackage{dcolumn}
\usepackage{bm}
\usepackage{braket}
\usepackage{fixltx2e}
\usepackage{amsmath}
\usepackage{siunitx}
\usepackage{changes}
\usepackage{natbib}
\usepackage[title]{appendix}

\usepackage{multirow}

\usepackage[utf8]{inputenc}

\usepackage[colorlinks]{hyperref}
\definecolor{blue}{RGB}{0,0,255}
\hypersetup{
urlcolor={blue}, 
citecolor={blue}, 
linkcolor={blue}, 
}

\begin{document}

\title{Operating in a deep underground facility improves the locking of gradiometric fluxonium qubits at the sweet spots}

\author{Daria~Gusenkova}
\affiliation{IQMT, Karlsruher Institute of Technology, Eggenstein-Leopoldshafen, Germany}
\affiliation{PHI, Karlsruhe Institute of Technology, Karlsruhe, Germany}

\author{Francesco~Valenti}
\affiliation{IQMT, Karlsruher Institute of Technology, Eggenstein-Leopoldshafen, Germany}

\author{Martin~Spiecker}
\affiliation{IQMT, Karlsruher Institute of Technology, Eggenstein-Leopoldshafen, Germany}
\affiliation{PHI, Karlsruhe Institute of Technology, Karlsruhe, Germany}

\author{Simon~G{\"u}nzler}
\affiliation{IQMT, Karlsruher Institute of Technology, Eggenstein-Leopoldshafen, Germany}

\author{Patrick~Paluch}
\affiliation{IQMT, Karlsruher Institute of Technology, Eggenstein-Leopoldshafen, Germany}
\affiliation{PHI, Karlsruhe Institute of Technology, Karlsruhe, Germany}

\author{Dennis~Rieger}
\affiliation{PHI, Karlsruhe Institute of Technology, Karlsruhe, Germany}

\author{Larisa-Milena Piora\c{s}-\c{T}imbolma\c{s}}
\affiliation{CETATEA, National Institute for Research and Development of Isotopic and Molecular Technologies, Cluj-Napoca, Romania}
\affiliation{Faculty of Physics, Babe\c{s}-Bolyai University, Cluj-Napoca, Romania}

\author{Liviu~P.~Z\^arbo}
\affiliation{CETATEA, National Institute for Research and Development of Isotopic and Molecular Technologies, Cluj-Napoca, Romania}

\author{Nicola~Casali}
\affiliation{INFN, Sezione di Roma,  Roma, Italy}

\author{Ivan~Colantoni}
\affiliation{INFN, Sezione di Roma,  Roma, Italy}
\affiliation{CNR, Istituto di Nanotecnologia, c/o Dip. Fisica, Sapienza Universit\`a di Roma, Roma, Italy}

\author{Angelo~Cruciani}
\affiliation{INFN, Sezione di Roma,  Roma, Italy}

\author{Stefano~Pirro}
\affiliation{INFN Laboratori Nazionali del Gran Sasso, Assergi (AQ), Italy}

\author{Laura~Cardani}
\affiliation{INFN, Sezione di Roma,  Roma, Italy}

\author{Alexandru~Petrescu}
\affiliation{Institut Quantique and D\'epartement de Physique, Universit\'e de Sherbrooke, Sherbrooke, Canada}

\author{Wolfgang~Wernsdorfer}
\affiliation{IQMT, Karlsruher Institute of Technology, Eggenstein-Leopoldshafen, Germany}
\affiliation{PHI, Karlsruhe Institute of Technology, Karlsruhe, Germany}

\author{Patrick~Winkel}
\affiliation{IQMT, Karlsruher Institute of Technology, Eggenstein-Leopoldshafen, Germany}

\author{Ioan~M.~Pop}
\email{ioan.pop@kit.edu}
\affiliation{IQMT, Karlsruher Institute of Technology, Eggenstein-Leopoldshafen, Germany}
\affiliation{PHI, Karlsruhe Institute of Technology, Karlsruhe, Germany}

\date{\today}

\begin{abstract}
We demonstrate flux-bias locking and operation of a gradiometric fluxonium artificial atom using two symmetric granular aluminum (grAl) loops to implement the superinductor. The gradiometric fluxonium shows two orders of magnitude suppression of sensitivity to homogeneous magnetic fields, which can be an asset for hybrid quantum systems requiring strong magnetic field biasing. By cooling down the device in an external magnetic field while crossing the metal-to-superconductor transition, the gradiometric fluxonium can be locked either at $0$ or $\Phi_0/2$ effective flux bias, corresponding to an even or odd number of trapped fluxons, respectively. At mK temperatures, the fluxon parity prepared during initialization survives to magnetic field bias exceeding $100 \,\Phi_0$. However, even for states biased in the vicinity of $1 \,\Phi_0$, we observe unexpectedly short fluxon lifetimes of a few hours, which cannot be explained by thermal or quantum phase slips. When operating in a deep-underground cryostat of the Gran Sasso laboratory, the fluxon lifetimes increase to days, indicating that ionizing events activate phase slips in the grAl superinductor.  
\end{abstract}

\maketitle


The unique properties of the superconducting state emerging in a select list of materials below the critical temperature $T_\mathrm{c}$ have already been used for particle detection~\cite{Zmuidzinas2004detectors, Irwin2005TES, Mazin2009MKID, Natarajan2012detectors}, quantum-limited amplification~\cite{Castellanos-Beltran2008, Bergeal2010amplification, ROY2016parametric}, quantum information processing~\cite{Devoret2013Outlook, Wendin_2017superconducting} and hybrid mesoscopic hardware~\cite{Xiang2013Hybrid}. While the main benefits offered by the superconducting state are unarguably its intrinsically low dissipation and the possibility to engineer strongly non-linear elements such as Josephson junctions (JJs), another potential resource emerges as a consequence of the magnetic field quantization in superconducting loops and the associated long-lived persistent currents~\cite{Kim1962Persistent, Little1967Persistent}. In classical superconducting circuits, trapped flux quanta called fluxons have been used for more than two decades in so-called rapid single flux-quantum electronics~\cite{Likharev1991RSFQ, Bunyk2001RSFQ}, and might constitute a valuable resource for local magnetic field biasing in superconducting quantum processors~\cite{Majer2002biasing}. In the quantum regime, fluxons have recently been proposed as a resource for quantum simulators~\cite{Shnirman2017Emulating, Petrescu2018Fluxon}.

A prominent example for fluxon dynamics is the fluxonium qubit, in which the tunneling of a fluxon through a JJ shunted by a superinductor~\cite{Bell2012Superinductor, Masluk2012superind, Peruzzo2020Oct} determines the eigenenergies and wavefunctions of the system \cite{manucharyan2009fluxonium}. Besides the large anharmonicity of its energy spectrum, enabling fast operation, the fluxonium exhibits a so-called sweet spot with a long energy relaxation time and slow dephasing when the magnetic flux enclosed in the loop is half a flux quantum~\cite{Pop2014quasiparticles, somoroff2021millisecond}. 

\begin{figure*}[t!]
\begin{center}
\includegraphics[width = 2\columnwidth]{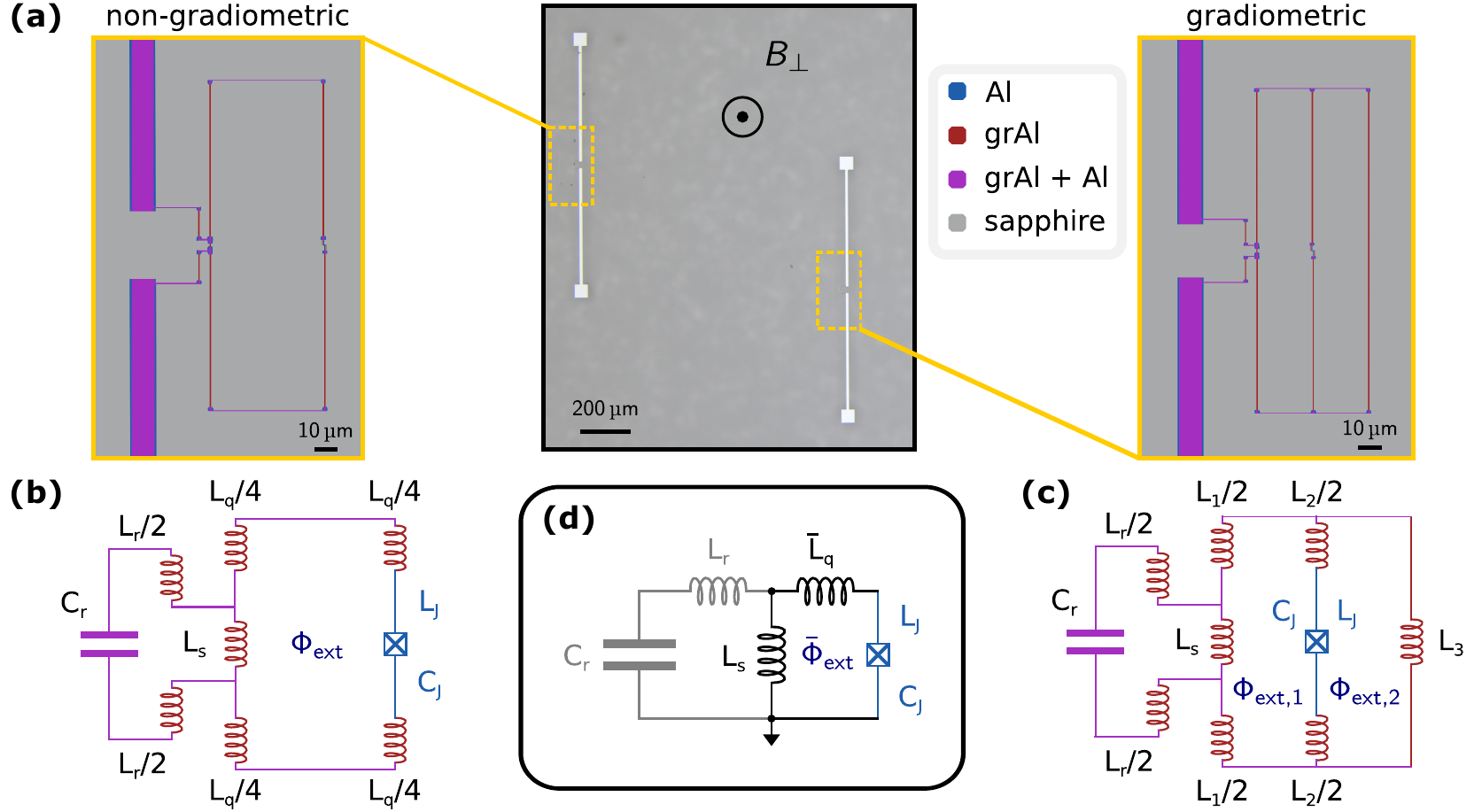}
\caption{\textbf{Gradiometric fluxonium.} \textbf{(a)} Optical microscopy images of our device, consisting of a pair of fluxonium artificial atoms (central panel) - a non-gradiometric fluxonium with a single loop (left panel) used to calibrate the external magnetic field, roughly $1\,\mathrm{mm}$ apart from a fluxonium with gradiometric loop (right panel). In both artificial atoms, a single JJ is shunted by a superinductor made out of grAl similar to Ref.~\cite{Grunhaupt2019Fluxonium}. The false colors indicate regions in which pure Al (blue), grAl (red), and a stack of both (purple) are deposited on a sapphire substrate (grey). For readout, the atoms are dispersively coupled to dedicated linear modes via a shared inductance. \textbf{(b)} Effective circuit diagram of the non-gradiometric fluxonium, where $C_\mathrm{r}$ and $L_\mathrm{r}$ are the capacitance and inductance associated with the linear readout mode, respectively, $L_\mathrm{s}$ is the shared inductance, $L_\mathrm{q}$ is the atom's loop inductance, and $L_\mathrm{J}$ and $C_\mathrm{J}$ are the Josephson inductance and capacitance, respectively. \textbf{(c)} Effective circuit diagram of the gradiometric fluxonium. The JJ is shunted by two superinductors, forming three loops in total. The flux enclosed in the two inner loops is $\Phi_\mathrm{ext,1}$ and $\Phi_\mathrm{ext,2}$. \textbf{(d)} Both implementations can be mapped onto a simplified circuit model. While the mapping of the non-gradiometric design is trivial, we find non-trivial expressions for the effective flux bias $\bar{\Phi}_\mathrm{ext}$ and the effective loop inductance $\bar{L}_\mathrm{q}$ for the gradiometric case (see supplementary material S1). }
\label{fig1}
\end{center}
\end{figure*} 

In our device, we use an additional superinductor made from a superconducting granular aluminum (grAl) \cite{Deutscher1973granular, Levy-Bertrand2019Electrodynamics} thin film shunting the single Al-AlOx-Al JJ to build a fluxonium artificial atom with a gradiometric loop geometry, as shown in Fig.\;\ref{fig1}. As a result, our device has three loops in total: an outer loop entirely formed by superinductors and two inner loops which are connected by a JJ weak link enabling quantum tunneling of fluxons between them. Similar to other gradiometric devices~\cite{Schwarz2013Gradiometric, Pita-Vidal2020Gate}, this loop geometry highly reduces the circuit's sensitivity to global magnetic fields, in our device by two orders of magnitude. This feature opens the way for its use in hybrid systems~\cite{Kubo11, Ranjan13} where a large magnetic field is required to bias other quantum degrees of freedom, for instance electronic spins in semiconducting heterostructures~\cite{Samkharadze18, Mi18} or molecular qubits~\cite{Bogani2008,Wernsdorfer19}.

The ground state for a superconducting loop threaded by a perpendicular external magnetic field involves a non-zero current, also known as persistent current, if the magnetic flux enclosed in the loop is not an integer multiple of the magnetic flux quantum $\Phi_0 = h / (2e)$. Similar to the Meissner effect, persistent currents can be induced by a static magnetic field if the superconducting loop is cooled below $T_\mathrm{c}$ and crosses the metal-to-superconductor phase transition. When the magnetic field is ramped down at temperatures well below $T_\mathrm{c}$, the changing magnetic field induces a screening current such that the number of trapped flux quanta in the loop remains constant~\cite{Masluk2012superind}. 

We demonstrate that our grAl gradiometric fluxonium can be initialized at the half-flux sweet spot by cooling down through $T_\mathrm{c}$ in a static magnetic field corresponding to $1 \,\Phi_0$ in the outer loop. From pulsed time-domain measurements we find an average energy relaxation time of $T_1 = 10.0 \pm 0.3\,\si{\micro\second}$ and a coherence time $T_2^\star = 0.59 \pm 0.02\, \si{\micro\second}$. Since the echo time $T_2 = 5.3 \pm \SI{0.3}{\micro\second}$ is roughly an order of magnitude larger, we infer that our device is limited by local low-frequency noise of unknown origin, qualitatively consistent with previous observations~\cite{Kou2017Molecule}.

Although the grAl superinductor is expected to have an extremely low phase-slip rate $\sim 10^{-20} \, \si{\hertz}$, we only observe a lifetime of the persistent current on the order of hours in a typical setup not shielded from ionizing radiation. The measured extinction of persistent current in our $50\times\SI{160}{\nano\meter}^2$ cross-section grAl wire is reminiscent of the operating principle of transition edge sensors~\cite{Irwin1995Apr} and superconducting nanowire single-photon detectors~\cite{Gol'tsman2001Aug,Korzh2020}, which, when DC current biased, can transition to a dissipative state due to a sudden burst of quasiparticles following an energy absorbing event. We confirm that the escape of trapped flux from the gradiometric loop is related to radioactivity by moving samples to the Gran~Sasso National Laboratory (LNGS) underground facility. Here, we measure a significant fluxon lifetime increase, from hours (above ground) to days. In the presence of a $\mathrm{ThO_2}$ radioactive source (same stup as in Ref.~\cite{Cardani2021underground}) this time reduces again to $\sim$30 minutes. 


\begin{figure*}[!t]
\begin{center}
\includegraphics[width = 2\columnwidth]{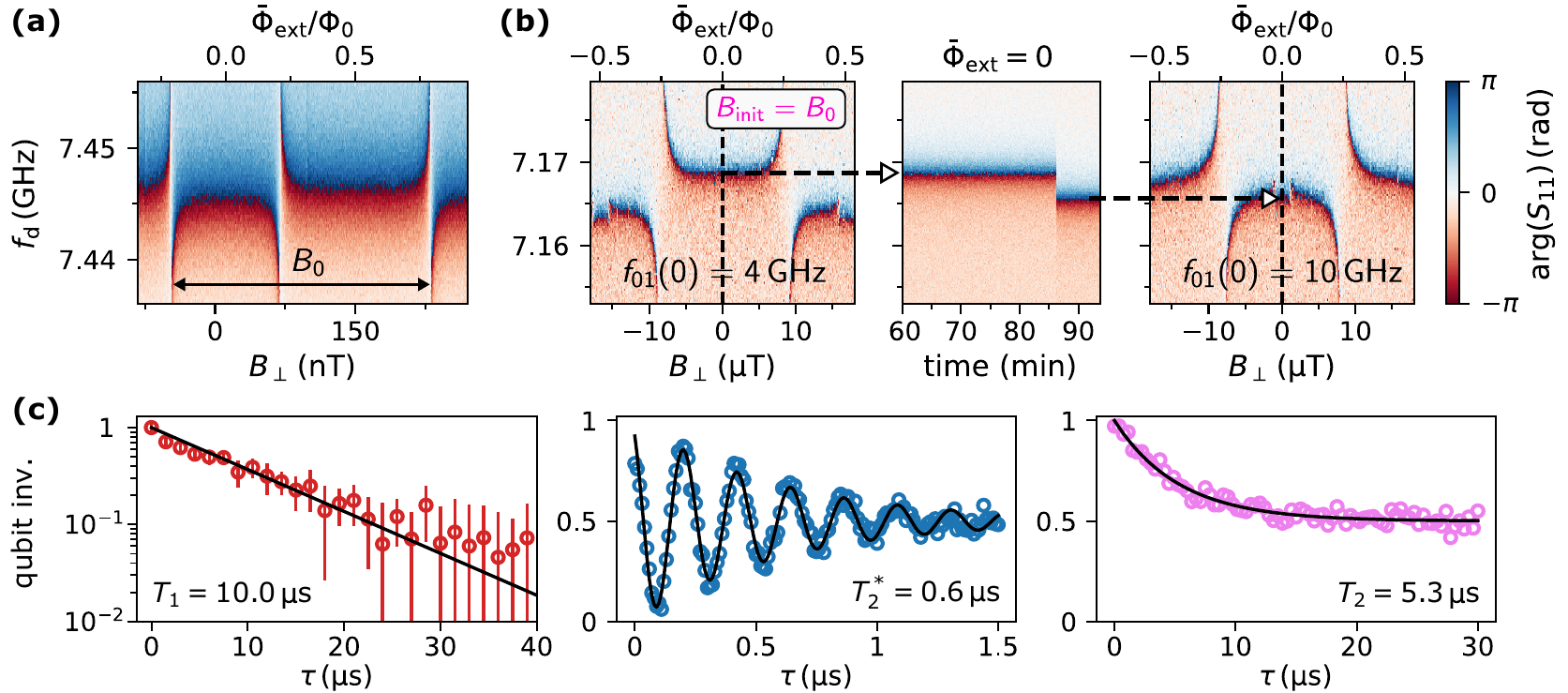}
\caption{\textbf{(a)} Calibration of the external field using the periodicity of the non-gradiometric fluxonium spectrum. The colorplot shows the phase of the reflection coefficient $\arg(S_{11})$ of the linear readout mode as a function of the external magnetic field $B_\perp$. The fundamental transition frequency of the fluxonium $f_{01}(\bar{\Phi}_\mathrm{ext})$ crosses the readout mode several times, resulting in repeated avoided crossings with a period $B_0 = \SI{280}{\nano\tesla}$ corresponding to a flux quantum $\Phi_0$ enclosed in the fluxonium loop. \textbf{(b)} Left panel: gradiometric fluxonium initialized at the effective half-flux bias by cooling down in $B_\mathrm{init} = B_0$. Notice the factor 120 reduced sensitivity of the gradiometric device to $B_\perp$ in comparison to panel (a). Central panel: the time trace of the phase response measured at $B_\perp=0$. The corresponding cut is indicated in left panel by a vertical dashed line. The jump of the frequency of the readout mode detected at $\approx85$~minutes after crossing $T_\mathrm{c,grAl} \approx 2\,\mathrm{K}$ corresponds to an escape of the trapped flux. Right panel: gradiometric device after the flux escape. The direction of the avoided crossings demonstrates that the fundamental fluxonium transition is found above (left) and below (right) the readout mode frequency in applied zero-field $B_\perp = 0$. The small avoided crossings visible in the vicinity of $B_\perp = 0$ in the right panel correspond to two-photon transitions. \textbf{(c)} Coherence of the gradiometric fluxonium after half-flux initialization: the qubit population inversion as function of time for energy relaxation (left), Ramsey fringes (center) and Hahn-echo experiment (right). Zero inversion corresponds to the finite population caused by thermal excitations at the fridge temperature of $\SI{20}{\milli\kelvin}$ and other non-equilibrium processes. The black lines indicate the numerical fit of the data (markers). Error bars in left panel show the measured standard deviation. } 
\label{fig2}
\end{center}
\end{figure*}

\begin{figure*}[!t]
\begin{center}
\includegraphics[width = 2\columnwidth]{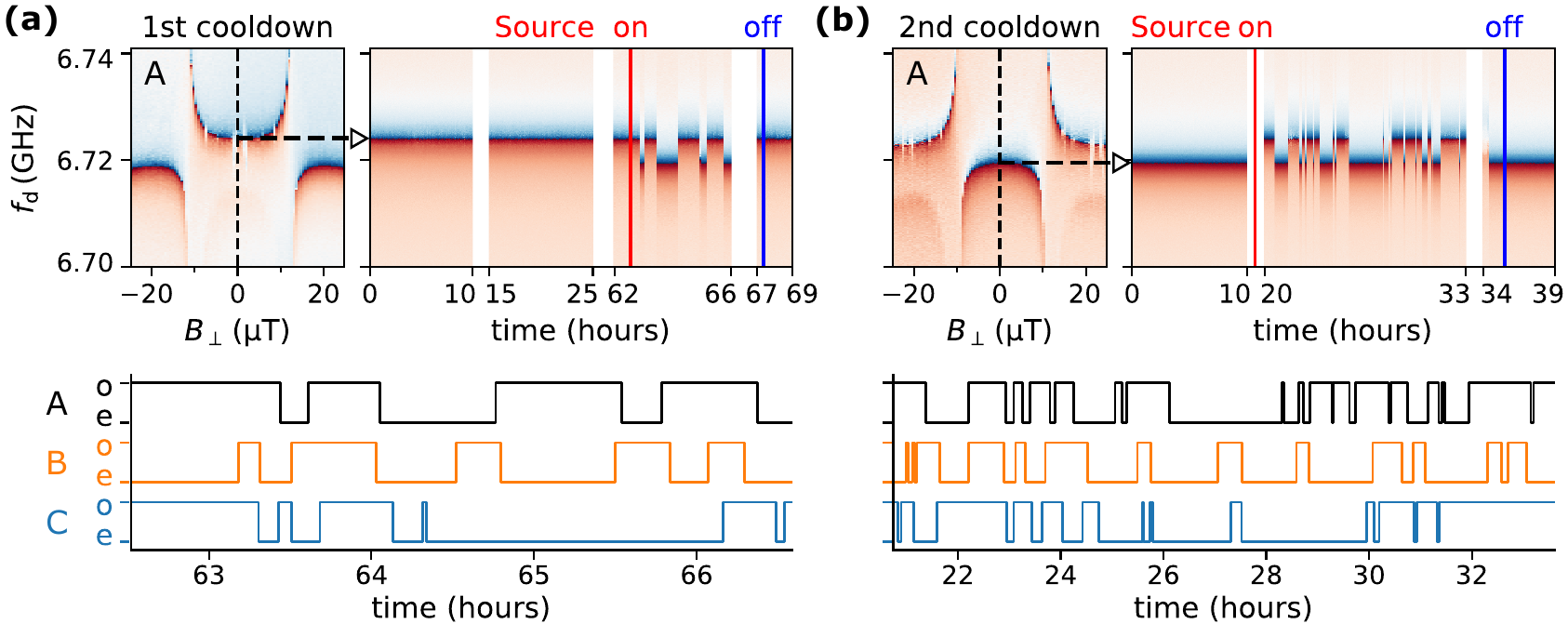}
\caption{Fluxon dynamics measured deep-underground in LNGS. The LNGS cryostat is located under a 1.4 km granite overburden (3.6 km water equivalent) and is additionally protected from ionizing radiation with lead shields located both inside and outside the refrigerator. We measured a chip with three gradiometric devices (labeled A, B and C) to check correlations between flux tunneling events. Top panels: the left-hand panels in (a) and (b) show the field dependence of device A in two separate cooldowns demonstrating odd and even state initialization, respectively. The right-hand panels show time traces measured at $B_\perp = 0$. Notice the stability of the trapped flux on timescales of days, before exposing the cryostat to a $\mathrm{ThO_2}$ radioactive source (red vertical line), which activates fluxon dynamics.  The blue vertical line indicates source removal. The bottom panels show measured switching dynamics between odd and even states for all devices during $\mathrm{ThO_2}$ exposure.}
\label{fig3}
\end{center}
\end{figure*}

The sample design, shown in Fig.\;\ref{fig1}\;(a), consists of a pair of fluxonium artificial atoms, one with a non-gradiometric and the other with a gradiometric loop geometry, respectively. Both devices are fabricated on a $0.33\times10\times\SI{15}{\milli\meter}^3$ c-plane sapphire substrate by means of a three-angle shadow evaporation, similar to Ref.~\cite{Grunhaupt2019Fluxonium} (see S2). The modulation periodicity of the non-gradiometric atom is used to calibrate the external magnetic flux created by a superconducting field coil. Although the devices are around $1\,\mathrm{mm}$ apart to reduce electromagnetic interaction, the diameter of the field coil is more than one order of magnitude larger, ensuring a homogeneous field $B_\perp$. For readout, both fluxonium atoms are dispersively coupled to dedicated readout modes by sharing a small fraction of their loop inductance. The capacitor of these two readout modes is designed in the form of a microwave antenna and couples them to the electric field of a 3D copper waveguide sample holder similar to Ref.~\cite{Grunhaupt2019Fluxonium}. 

For both device geometries we derive effective lumped-element circuit models (see Fig.\;\ref{fig1} panels (b) and (c)). Since the readout is implemented similarly, the capacitance and inductance of the readout modes are denoted $C_\mathrm{r}$ and $L_\mathrm{r}$, respectively, and $L_\mathrm{s}$ is the shared inductance. The non-gradiometric design has a single loop with a superinductance $L_\mathrm{q}$ shunting the JJ (blue crossed-box symbol). The gradiometric design has two shunt inductances forming three loops: an outer loop with surface area $A=50\times\SI{150}{\micro\meter}^2$, and two inner loops with surface area $A / 2$. The inductance in each loop branch is denoted $L_i$, with the index $i \in \{1,2,3\}$ indicating the corresponding branch. The  gradiometric atom can be mapped onto the standard fluxonium circuit diagram shown in Fig.\;\ref{fig1}\;(d) using an effective flux bias $\bar{\Phi}_\mathrm{ext}$ and an effective shunting inductance $\bar{L}_\mathrm{q}$ (see S1). 

The superconducting field coil is calibrated by measuring the spectrum of the non-gradiometric device, designed with the same loop area $A$, located on the same substrate. Figure~\ref{fig2}\;(a) depicts the phase response $\arg(S_{11})$ of the readout mode coupled to the non-gradiometric fluxonium atom as a function of the probe frequency $f_\mathrm{d}$ and the external magnetic field $B_\perp$, measured in close vicinity of the readout frequency $f_\mathrm{r} = 7.445\,\mathrm{GHz}$. The fundamental transition frequency of the fluxonium $f_{01}(\bar{\Phi}_\mathrm{ext})$ oscillates between values below and above the readout frequency, resulting in avoided-level-crossings repeated with periodicity of $B_0 = \SI{0.28}{\micro\tesla}$. 


The gradiometric fluxonium can be initialized at the half-flux effective bias by cooling the device down through the metal-to-superconductor phase transition in a static magnetic field $B_\mathrm{init} = B_0$ corresponding to a single flux quantum enclosed in the outer fluxonium loop (see S3). The magnetic field is ramped down at the base temperature of the cryogenic refrigerator ($\SI{20}{\milli\kelvin}$), well below the critical temperature $T_\mathrm{c,grAl} \approx 2\,\mathrm{K}$ of the grAl film. However, the enclosed flux is now trapped in the gradiometric loop. In case of perfectly symmetric inner loops and zero field gradient the phase difference across the JJ equals $\pi$, pinning the atom at the half-flux bias. Fig.\;\ref{fig2}\;(b) shows the gradiometric fluxonium after initialization at the effective half-flux bias (left panel). Wide range flux sweeps of the gradiometric device are shown in S5. The difference in field range covered in  Fig.\;\ref{fig2}\;(a) and Fig.\;\ref{fig2}\;(b)  illustrates the suppression of global magnetic field sensitivity by roughly a factor of 120 for the gradiometric fluxonium. According to our effective circuit model, the remaining field sensitivity could be either caused by an asymmetry of the outer loop inductances, or by a small field gradient.

Figure~\ref{fig2}\;(c) depicts time-domain characterization of the coherence properties of the gradiometric atom. For the gradiometric fluxonium initialized at the effective half-flux bias we find a Ramsey coherence time of $T_2^\star = 0.59 \pm 0.02\, \si{\micro\second}$, which is not limited by the energy relaxation time $T_1 = 10.0 \pm 0.3 \,\si{\micro\second}$. We measured $T_1$ fluctuations of $10\%$ on a timescale of two hours.  Notably, the non-gradiometric fluxonium located on the same chip exhibits similar coherence times $T_1 = 2.5 \pm 0.3\,\si{\micro\second}$ and $T_2^\star = 0.76 \pm 0.04\, \si{\micro\second}$, which excludes the gradiometric geometry as the cause of the much smaller coherence compared to previous fluxonium implementations based on similar grAl superinductors \cite{Grunhaupt2019Fluxonium}. Moreover, in both devices we do not observe an improvement in coherence around the half-flux sweet spot (see S4). While the sensitivity to homogeneous fields is decreased for the gradiometric device, this is not the case for local flux noise, which might even increase due to larger length of the shunting inductance~\cite{Lanting2009}.  A single spin echo pulse improves the coherence by almost an order of magnitude for the gradiometric fluxonium, up to $T_2 = 5.3 \pm \SI{0.3}{\micro\second}$, and by factor of 3.5 for the non-gradiometric fluxonium, up to $T_2 = 2.6 \pm \SI{0.4}{\micro\second}$.  Therefore, we conclude that Ramsey coherence of all devices on this chip is limited by local and low-frequency noise of unknown origin.

The time stability of the half-flux initialization is determined by fluxon escape rate, which becomes apparent by an abrupt change of persistent current under constant or zero magnetic field bias. To suppress fluxon dynamics the outer loop of gradiometric devices needs to be implemented using a superconducting wire with low phase slip rate. The expected phase slip rate in our grAl superinductance can be found by modeling the material as an effective array of JJs \cite{Maleeva2018Circuit}. The calculated phase-slip rate is $\sim 10 ^{-20} \, \SI{}{\hertz}$ (see S5). In strong contrast, in all four cooldowns in the cryostat located in Karlsruhe (not shielded from ionizing radiation) we observe an escape of the trapped flux once in a few hours, similar to the phase slip rate found in conventional JJ array  superinductors~\cite{Masluk2012superind}. The time evolution of the readout mode in Fig.\;\ref{fig2}\;(b) shows a detected flux escape event, manifesting as a frequency jump at $\approx85$~minutes after crossing $T_\mathrm{c,grAl}$. In order to test whether these jumps are caused by ionizing radiation~\cite{Swenson2010phonon, Grunhaupt2018Quasiparticle, Vepsalainen2020Aug, McEwen2021Apr, Wilen2021Correlated} 
we measure three similar gradiometric devices in the LNGS deep-underground facility (Fig.\;\ref{fig3}), which was previously used to quantify non-equilibrium quasiparticle poisoning in superconducting quantum circuits~\cite{Cardani2021underground}. For all  devices measured in LNGS the trapped flux remains stable on a time scale of days. Exposing the cryostat to a $\mathrm{ThO_2}$ radioactive source leads to uncorrelated flux tunneling events, and reduces fluxon lifetime to approximately half an hour. The fluxon  stability is restored after removal of the source.


In summary, we have demonstrated the implementation of a superconducting fluxonium artificial atom with a gradiometric loop geometry, which is two orders of magnitude less sensitive to global magnetic fields compared to a standard, non-gradiometric device with similar loop area. We can initialize our device at the half-flux sweet spot by inducing a persistent current into the outer loop when cooling it down through the metal-to-superconductor transition in a static external magnetic field of $B_\mathrm{init} = \SI{0.28}{\micro\tesla}$, equivalent to a single flux quantum enclosed in the outer loop. From pulsed time-domain measurements we find that the coherence of the gradiometric device is comparable to that of regular fluxoniums on the same chip, and it is limited by local, low-frequency noise, which can be filtered by a single spin echo pulse. 

Although the superinductor in our device is implemented using superconducting grAl, which is expected to have a significantly smaller phase-slip rate compared to conventional JJ arrays~\cite{Masluk2012superind}, we observe a similar escape rate of the flux after half-flux initialization, which is indicative of catastrophic events, for instance caused by radioactive or cosmic impacts locally weakening superconductivity in the outer loop wire. Indeed, we confirm this hypothesis by measuring order of magnitude increased lifetimes of trapped fluxons in the LNGS deep-underground facility. Our results add another item to the list of detrimental effects of ionizing radiation in superconducting hardware, and provide additional motivation to implement radiation mitigation~\cite{Karatsu2019Jan, Henriques2019traps, Cardani2021underground, Martinis2021saving, McEwen2021Apr}.

See supplementary material for the Hamiltonian of gradiometric fluxonium, sample fabrication, gradiometric fluxonium initialization at half flux bias, measured spectrum and coherence of the gradiometric fluxonium, and escape of the trapped flux from grAl loop. 


We are grateful to L.~Gr{\"u}nhaupt for fruitful discussions, and we acknowledge technical support from S.~Diewald, A.~Lukashenko and L.~Radtke. Funding is provided by the Alexander von Humboldt foundation in the framework of a Sofja Kovalevskaja award endowed by the German Federal Ministry of Education and Research, and by the European Union's Horizon 2020 programme under Nº 899561 (AVaQus). P.P. and I.M.P. acknowledge support from the German Ministry of Education and Research (BMBF) within the QUANTERA project SiUCs (FKZ: 13N15209). L.P.Z and L-M. P-\c{T}. acknowledge the Romanian National Authority for Scientific Research and Innovation, CNCS-UEFISCDI for funding through the project PCCF Project No. PN-III-P4-ID-PCCF-2016-0047 and the project Quantum Computation with Schr{\"o}dinger cat states, contract ERANET-QUANTERA-QuCos 120/16.09.2019.  S.G., D.R. and W.W. acknowledge support from the European Research Council advanced grant MoQuOS (N. 741276). Facilities use was supported by the KIT Nanostructure  Service  Laboratory  (NSL). We acknowledge qKit for providing a convenient measurement software framework. We thank the director and technical staff of INFN-LNGS and the LNGS Computing and Network Service for their support. We are grateful to M.~Guetti and M.~Iannone for the support with construction of the cryogenic setup at LNGS. The work was supported by INFN and by the U.S. Department of Energy, Office of Science, National Quantum Information Science Research Centers, Superconducting Quantum Materials and Systems Center (SQMS) under the contract No. DE-AC02-07CH11359.

The data that support the findings of this study are available from the corresponding author upon reasonable request.

\hfill

\bibliography{main}

\onecolumngrid
\vspace{5cm}
\clearpage
\section*{Supplementary Material}

\renewcommand{\thesubsection}{S\arabic{subsection}}
\renewcommand{\thetable}{S\arabic{table}}
\renewcommand{\thefigure}{S\arabic{figure}}
\counterwithin{figure}{subsection}
\counterwithin{equation}{subsection}

In this supplementary material we provide details on the Hamiltonian of gradiometric fluxonium, sample fabrication, the initialization of the gradiometric fluxonium at the half-flux bias point, measured spectrum and coherence of the gradiometric fluxonium, and escape of the trapped flux from grAl loop.
\vspace{1.0cm}
\twocolumngrid

\subsection{Fluxonium equivalent circuit model}

Using the lumped-element circuit in Fig.\;1\;(c) in the main text as a starting point for obtaining a quantum theory, we arrive at the following Hamiltonian describing readout resonator, gradiometric fluxonium qubit, and their linear inductive coupling:
\begin{align}
  \begin{split}
  \boldsymbol{H} = & \frac{1}{2C_\mathrm{r}}  \boldsymbol{Q}_\mathrm{r}^2 + 
    \frac{1}{2 \bar{L}_\mathrm{r}} \boldsymbol{\Phi}_\mathrm{r}^2  \\ &+  \frac{1}{2C_\mathrm{q}}  \boldsymbol{Q}_\mathrm{q}^2  +\frac{1}{2 \bar{L}_\mathrm{q}} \boldsymbol{\Phi}_\mathrm{q}^2 - E_\mathrm{J} \cos\left(\boldsymbol{\varphi}_\mathrm{q} + \frac{2\pi}{\Phi_0} \bar{\Phi}_\text{ext}\right) \\ &-\frac{1}{2 \bar{L}_{\mathrm{rq}}}\boldsymbol{\Phi}_\mathrm{r} \boldsymbol{\Phi}_\mathrm{q} , \label{Eq:H}
  \end{split}
\end{align}
Here, we have introduced two pairs of canonically-conjugate charge and flux variables, $\left[ \boldsymbol{\Phi}_i , \boldsymbol{Q}_j \right] = i \hbar \delta_{ij}$ for $i,j = q,r$ corresponding to the qubit and resonator, respectively. For the qubit, we defined the superconducting phase $\boldsymbol{\varphi}_\mathrm{q} = 2\pi \boldsymbol{\Phi}_\mathrm{q}/\Phi_0$, with $\Phi_0$ the superconducting flux quantum.

\begin{figure}[t!]{}
  \includegraphics[width=\linewidth]{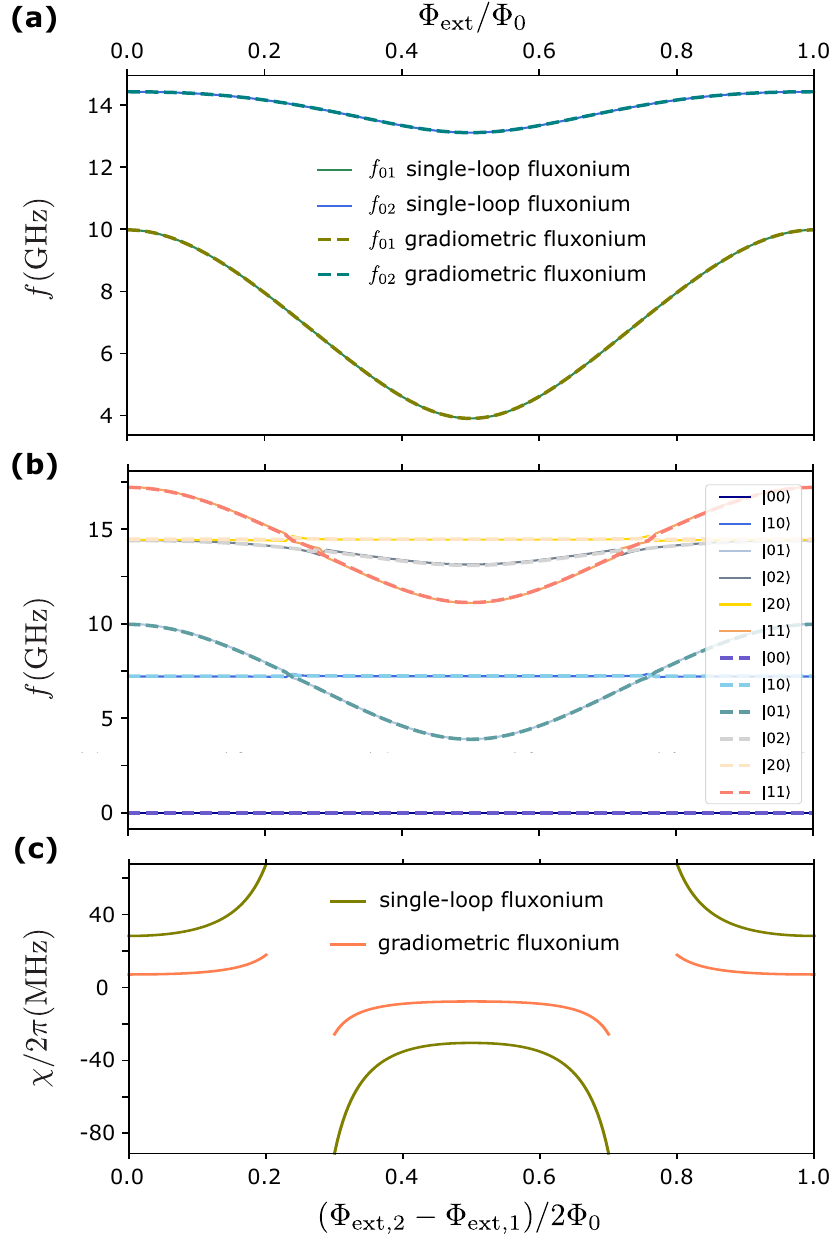}
  \caption{Comparison between the spectra corresponding to a single-loop fluxonium qubit and the gradiometric fluxonium, for the parameters stated in the main text. \textbf{(a)} The fluxonium spectrum as function of the external flux $\Phi_{\text{ext}}$ (solid lines) maps to that of a gradiometric fluxonium qubit (dashed lines) as a function of the relative flux $(\Phi_{\text{ext,2}}-\Phi_{\text{ext,1}})/2$ in the two smaller loops. \textbf{(b)}  Lowest energy levels $\left|n_\mathrm{r} m_\mathrm{q} \right\rangle$ in the combined spectra of the qubit-resonator system. The transitions primarily associated to the linear readout mode are bias field independent and visible as horizontal lines. Thanks to the finite shared inductance, avoided-level-crossings are visible whenever a fluxonium transition crosses the readout mode.  \textbf{(c)}  Dispersive shift $\chi$ as extracted from the curves shown in panel (b); excluded are areas near avoided crossings where a proper distinction between linear readout mode and non-linear qubit mode becomes ambiguous. \label{FigureS1}}
\end{figure}

The gradiometric circuit is defined in terms of effective fluxonium shunt, resonator, and coupling inductances
\begin{align}
\begin{split}
  \bar{L}_\mathrm{q} &= \left[\dfrac{L_3 L_\sigma^2(L_s+L_\mathrm{r})+L_\mathrm{r} L_s L_2 L_3}{L_\sigma^2 (L_\mathrm{r} L_a^2+L_s L_b^2)}+\frac{L_1}{L_\sigma^2}\right]^{-1},  \label{Eq:Lq} \\
  \bar{L}_\mathrm{r} &= \frac{L_\mathrm{r} L_a^2 +L_s L_b^2}{L_\epsilon^2}, \\
  \bar{L}_{\text{rq}} &= \frac{L_\mathrm{r} L_a^2 + L_s L_b^2 }{2 L_s L_3},
\end{split}
\end{align}
where $L_\sigma^2  = L_1 L_2 +L_2 L_3 +L_1 L_3$, $L_\epsilon^2 =  L_s L_2 +L_s L_3 + L_\sigma^2$, $L_a^2 = L_s L_2 + L_\sigma^2$, and $L_b^2 = L_\mathrm{r} L_3 + L_\sigma^2$. In the system Hamiltonian $\boldsymbol{H}$, the external fluxes enter via the linear combination 
\begin{align}
  \begin{split}
    \bar{\Phi}_\text{ext} 
        =& \Phi_{\Delta}+ \alpha \Phi_\Sigma, \; \alpha = \frac{L_3 - L_1 - L_s}{L_1+L_s+L_3}
  \end{split}
\end{align}
where $\Phi_{\Sigma/\Delta} = (\Phi_{\text{ext},1} \pm \Phi_{\text{ext},2})/2$, and $\alpha$ is the inductance asymmetry. A symmetric configuration, $\alpha=0$, leads to insensitivity to variations in the global magnetic flux. Nevertheless, the spectrum always depends on the flux imbalance $\Phi_\Delta$. 

\setcounter{subsection}{3}
\setcounter{figure}{0}   
\begin{figure*}[t!]
\begin{center}
\includegraphics[width = 2\columnwidth]{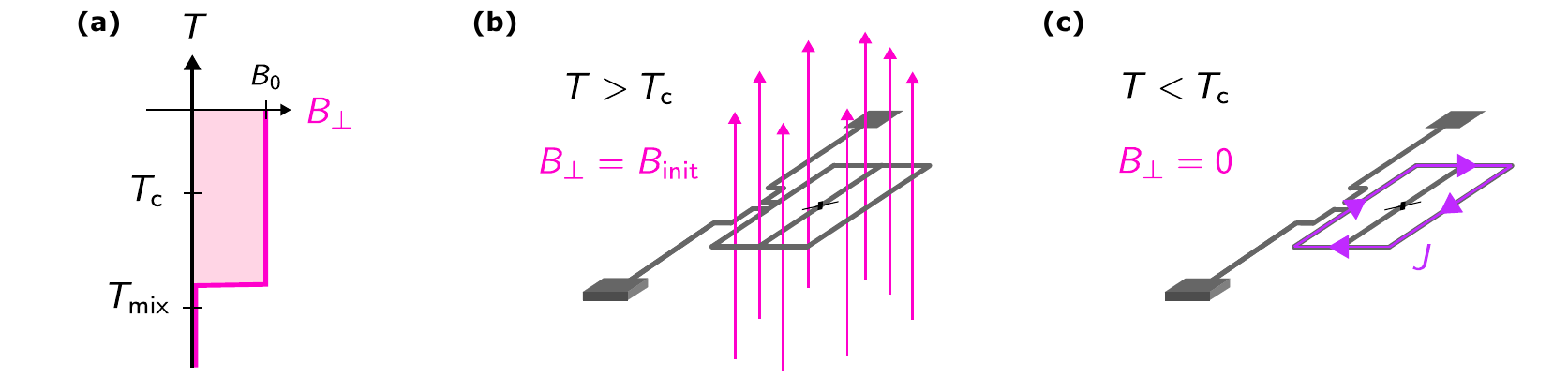}
\caption{\textbf{Flux bias initialization sequence.} \textbf{(a)} The initialization of the gradiometric fluxonium at the effective half-flux bias is realised by cooling down the device through the metal-to-superconductor phase transition in an external magnetic field $B_\mathrm{init} = B_0$ corresponding to single flux quantum in the outer fluxonium loop. Magnetic field is ramped down at the base temperature of the cryogenic refrigerator mixing chamber plate ($T_\mathrm{mix} = \SI{20}{\milli\kelvin}$), well below the critical temperature of grAl $T_\mathrm{c} \approx 2\,\mathrm{K}$ of the grAl film. \textbf{(b)} Schematic drawing of the external flux bias above $T_\mathrm{c}$. \textbf{(c)} Circulating current (purple arrow) induced after ramping down external magnetic field corresponds to the effective half-flux bias $\bar{\Phi}_\mathrm{ext} = 0.5 \Phi_0$. }
\label{FigureS3}
\end{center}
\end{figure*} 
\setcounter{subsection}{1}

As noted before in the main text, the qubit Hamiltonian in eq.~\ref{Eq:H} can be made exactly equivalent to the Hamiltonian of a single-loop fluxonium. Explicitly, setting the shared inductance $L_s=0$ to decouple the qubit from the resonator,  and putting further $L_2 = 0$ and $L_1 = L_3 = 2 L_q$, we arrive at the Hamiltonian of a single-loop fluxonium qubit with effective flux equal to the flux imbalance $\Phi_\Delta$ and shunt inductance $\bar{L}_\mathrm{q} = L_q$. In Fig.\;\ref{FigureS1}\;(a), we exemplify this exact agreement for the effective fluxonium parameters obtained in the main text.

Turning our attention to the circuit QED setup at $L_s \neq 0$, to numerically diagonalize the two-mode Hamiltonian $\boldsymbol{H}$, we recast it into the Fock basis determined by the terms \cite{Smith2016numerics} in the first two rows of eq.~\ref{Eq:H}. The lowest energy states in the spectrum of the combined qubit-resonator Hamiltonian $\left|n_\mathrm{r} m_\mathrm{q} \right\rangle$ and the resonator dispersive shift $\chi$ are shown in Fig.\;\ref{FigureS1}, panels (b) and (c), respectively. For the sweep over external flux for the qubit-resonator Hamiltonian, we have employed $m=25$ Fock states for the qubit, and $n=15$ states for the resonator mode, and have used larger-scale simulations of up to a total Hilbert space dimension of 3500 in order to obtain a numerically-precise value for the dispersive shift at the half-flux sweet spot, $\chi/2\pi = - 7.683\, \mathrm{MHz}$, which is in good agreement with the experimentally obtained shift $\chi_\mathrm{exp}/2\pi = - 7.8\, \mathrm{MHz}$.

\subsection{Fabrication}

We use a three-angle shadow evaporation technique \cite{Grunhaupt2019Fluxonium} to fabricate our fluxonium devices on a c-plane sapphire substrate. First, a $20\,\mathrm{nm}$ thick Al film is evaporated under an angle of $+30^\circ$, measured from the normal, which is then oxidized for 340~seconds under an oxygen partial pressure of $\SI{30}{\milli \bar}$. Second, another $30\,\mathrm{nm}$ thick Al film is evaporated under an angle $-30^\circ$, overlapping with the first layer and forming the JJ, followed by a $50\,\mathrm{nm}$ thick grAl film deposited at zero-angle. Thanks to the undercut in the double resist stack, most of the grAl loop wire remains unshunted by the pure Al films, while the JJ is formed entirely between the two Al films.

\subsection{Gradiometric fluxonium initialization at half flux bias}

The initialization sequence of the gradiometric fluxonium at the half-flux bias point is illustrated in Fig.\;\ref{FigureS3}. First, the system is cooled down through the metal-to-superconductor phase transition in an external magnetic field $B_0$. Thanks to the diameter of the superconducting field coil used in this experiment, the magnetic field can be approximated as homogeneous on the length-scale of our device. Notably, prior to this sequence the external magnetic bias field $B_0$ is calibrated using a non-gradiometric device on the same wafer (see Fig.\;1 and Fig.\;2 in the main text). Well below the critical temperature of our grAl film, which is on the order of $2\,\mathrm{K}$, the external magnetic field is ramped down, which induces a persistent current $J$ in the outer loop, indicated by the purple arrows in Fig.\;\ref{FigureS3}\;(c). The total flux enclosed in the loop corresponds to a single flux quantum.  

\subsection{Spectrum and coherence of the gradiometric fluxonium}

\begin{figure*}[t!]
\begin{center}
\includegraphics[width = 2\columnwidth]{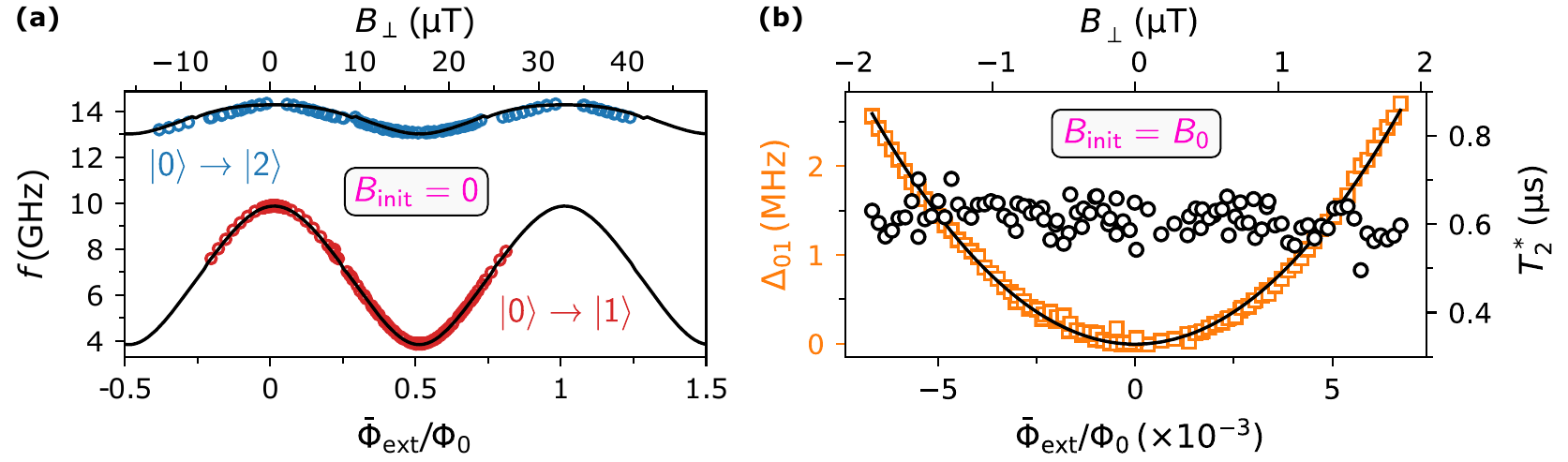}
\caption{\textbf{Spectrum and coherence of the gradiometric fluxonium.}  \textbf{(a)} Spectrum of the gradiometric fluxonium measured as a function of the effective magnetic flux $\bar{\Phi}_\mathrm{ext}$ and the applied magnetic field $B_\perp$, when initialized in zero-field $B_\mathrm{init} = 0$. The red markers indicate the fundamental transition frequency $f_{01}$ extracted from a continuous-wave two-tone measurement, by monitoring the readout resonator response while sweeping the frequency of an additional drive tone. The blue markers correspond to transition to the second excited state. The black solid lines indicate a fit to the fluxonium spectrum obtained by diagonalizing the uncoupled fluxonium Hamiltonian numerically. From the fit, we extract the effective circuit parameters $\bar{L}_\mathrm{q} = 172\,\mathrm{nH}$, $C_\mathrm{J} = 3.4\,\mathrm{fF}$, and $E_\mathrm{J} / h = 5.1 \, \mathrm{GHz}$. Notably, since we change the spectrum only by applying a global magnetic field, the value of the magnetic field at which we reach the half-flux sweet spot corresponds to more than $50 \, \Phi_0$ enclosed in the outer loop.  \textbf{(b)} Ramsey coherence time $T_2^\star$ (black markers) measured around spectrum minimum for the gradiometric fluxonium initialized at the half-flux bias ($B_\mathrm{init} = B_0$).  Notice that after initialization the minimum of the qubit fundamental frequency corresponds to zero applied field. \textcolor{black}{Orange markers show the detuning of the qubit fundamental transition frequency $\Delta_{01} = f_{01}(\bar{\Phi}_\mathrm{ext}) - f_{01}(\bar{\Phi}_\mathrm{ext}=0)$ from its frequency at the degeneracy point $f_{01}(\bar{\Phi}_\mathrm{ext}=0)$.} The black line is a parabolic fit. }
\label{FigureS41}
\end{center}
\end{figure*}

\setcounter{subsection}{5}
\setcounter{figure}{0} 
\begin{figure*}[t!]
\begin{center}
\includegraphics[width = 2\columnwidth]{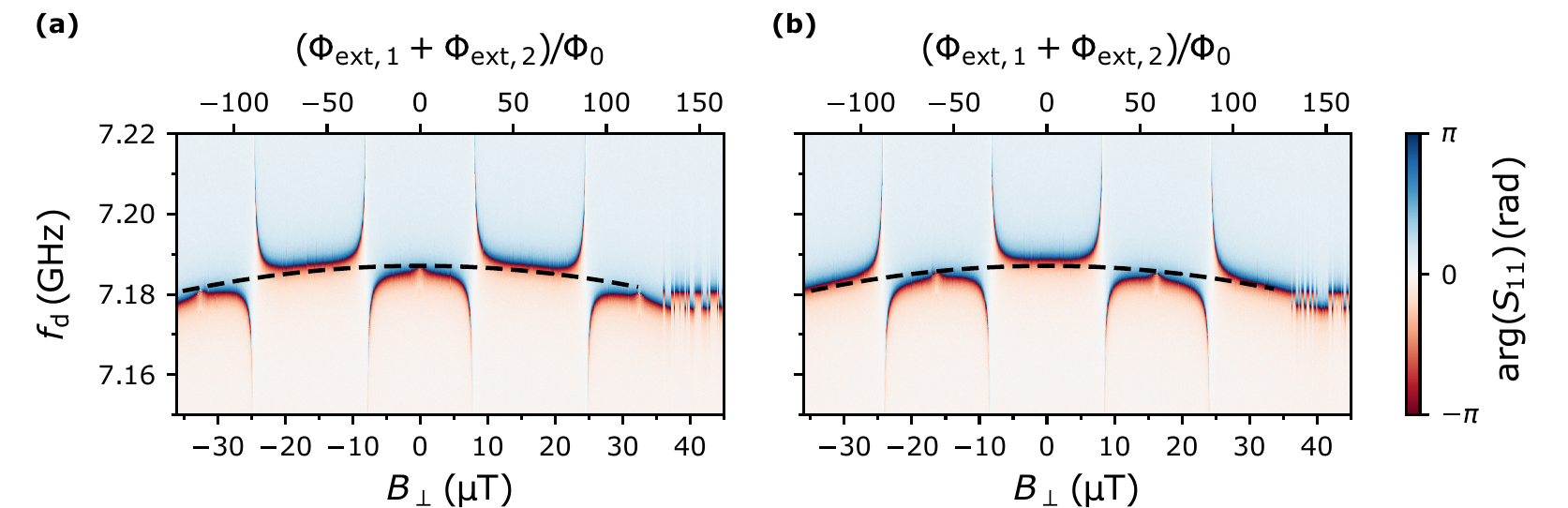}
\caption{\textbf{Current-activated phase slips in the gradiometric fluxonium.} Color plots show wide flux sweeps of the phase response of the readout mode coupled to the gradiometric fluxonium with even \textbf{(a)} and odd \textbf{(b)} flux quanta trapped in the grAl loop. The parabolic change of the frequency (black dashed lines) is due to the dependence of the grAl kinetic inductance on the circulating current induced by applied magnetic field in the fluxonium loop. This circulating current also flows through the coupling segment of the readout resonator. The top axes shows the applied magnetic field in units of flux quantum in the fluxonium outer loop. The jumps of frequency observed at $\gtrsim 130 \Phi_0$ correspond to current-activated phase slips. One flux sweep takes approximately 5 minutes.}
\label{FigureS51}
\end{center}
\end{figure*}
\setcounter{subsection}{4}

We measure the spectrum of the gradiometric fluxonium using two-tone spectroscopy (see Fig.\;\ref{FigureS41}\;(a)), meaning we continuously probe the readout resonator with a weak tone on resonance while applying a second tone with varying frequency to the qubit. In the investigated frequency range between $3\,\mathrm{GHz}$ and $15\,\mathrm{GHz}$ we find two field dependent transitions associated with the gradiometric fluxonium, a single photon transition into the first excited state $0 \rightarrow 1$ and the second excited state $0 \rightarrow 2$ shown in red and blue, respectively. The fluxonium circuit parameters $\bar{L}_\mathrm{q} = 172\,\mathrm{nH}$, $C_\mathrm{J} = 3.4\,\mathrm{fF}$, and $E_\mathrm{J} / h = 5.1 \, \mathrm{GHz}$ are extracted by fitting the spectrum to the numerically diagonalized fluxonium Hamiltonian. For the linear readout mode we find $C_\mathrm{r} = \SI{20.2}{\femto \farad}$ and $L_\mathrm{r} = \SI{21.6}{\nano\henry}$. The value of the shared inductance $L_\mathrm{s} = \SI{2.8}{\nano\henry}$ is extracted from the dispersive shift {$\chi_\mathrm{ge} (0.5 \Phi_0) / (2 \pi) = - 7.8 \,\mathrm{MHz}$} measured at the half-flux sweet spot. 

Fig.\;\ref{FigureS41}\;(b) shows the Ramsey coherence time $T_2^\star$ of the gradiometric fluxonium measured after the half-flux bias initialization around the spectrum minimum (black markers). \textcolor{black}{The detuning of the  fundamental transition frequency of the gradiometric fluxonium is shown in orange markers.} In contrast to previous fluxonium implementations \cite{Grunhaupt2019Fluxonium, Pop2014quasiparticles}, the device does not exhibit an improvement in coherence around the half-flux sweet spot. 

\subsection{Escape of the trapped flux from grAl loop}

Quantum phase slips of the superconducting phase in the granular aluminum superinductor are a potential decay mechanism for the trapped flux. The expected phase slip rate can be calculated by modeling grAl superinductor as an effective array of JJs \cite{Maleeva2018Circuit}.  GrAl consists of pure Al grains embedded in an insulating $\mathrm{AlO_x}$ matrix. With a typical grain size of $\sim \SI{4}{\nano\meter}$ we find the number of effective junctions in the outer loop $N \approx 75 \times 10^3$, with $E_\mathrm{J} = (\Phi_0/2\pi)^2 / L_\mathrm{J0} \approx \SI{53}{\tera \hertz}$ and $E_\mathrm{C} = e^2 / 2 C_\mathrm{J} \approx \SI{48}{\giga\hertz}$. The phase slip rate in the Josephson junction array~\cite{Pop2010slips, Rastelli2013QPS} is
\begin{equation}
    v = \frac{1}{h} N  \frac{4}{\sqrt{\pi}} (8 E_\mathrm{J}^3 E_\mathrm{C})^{1/4} 
    \exp \left( -  \sqrt{8\frac{E_\mathrm{J}}{E_\mathrm{C}}}\right),
    \label{eq:phase_slip}
\end{equation} 
which gives $v \lesssim 10 ^{-20} \, \SI{}{\hertz}$ for the grAl superinductor in our device. We would like to point out that the expected rate is orders of magnitude smaller compared to conventional JJ arrays~\cite{Masluk2012superind}. \textcolor{black}{The phase slip rate is enhanced by the induction of circulating currents because the effective Josephson barrier between the grAl grains is reduced. For the gradiometric fluxonium measured in a wide flux range we observe current-activated phase slips at an external magnetic field bias corresponding to $\gtrsim 130 \Phi_0$ on the outer loop (see Fig.\;\ref{FigureS51}). The fact that we can prepare such a highly metastable state in the fluxonium loop confirms the stiffness of the condensate in grAl.}

\end{document}